\documentclass[11pt]{article}
\usepackage{fullpage}
\usepackage{mathtools}
\usepackage[margin = 2.5cm]{geometry}

\usepackage{amsmath, amssymb, amsthm}

\usepackage{graphicx}
\usepackage{color}
\usepackage[dvipsnames]{xcolor}

\usepackage{pgfplots}
\pgfplotsset{compat=1.15}

\usepackage{tikz}
\usepackage[hyperindex,breaklinks]{hyperref}

\usepackage{mdframed}

\usepackage[shortlabels]{enumitem}

\usepackage[utf8]{inputenc} 
\usepackage[T1]{fontenc}    

\usepackage{amsmath,amsthm,amsfonts,bbm}

\usepackage[round]{natbib}
\usepackage{cleveref}

\usepackage{pgfplots}
\pgfplotsset{compat=1.15}

\usepackage{tikz}
\usepackage{graphicx}
\usepackage{caption}

\usepackage{algorithm}
\usepackage{algorithmic}

\usepackage{cleveref}
\usepackage{hyperref} 
\usepackage{url}

\usepackage{mathtools}
\usepackage{booktabs}
\usepackage{microtype}

\usepackage{xfrac}
\usepackage{nicefrac}    

\usepackage{dsfont}

\newtheorem{theorem}{Theorem}[section]
\newtheorem*{theorem*}{Theorem}

\newtheorem*{proposition*}{Proposition}
\newtheorem{lemma}[theorem]{Lemma}
\newtheorem*{lemma*}{Lemma}
\newtheorem{corollary}[theorem]{Corollary}
\newtheorem*{conjecture*}{Conjecture}

\newtheorem*{fact*}{Fact}

\newtheorem*{hypothesis*}{Hypothesis}

\newtheorem*{claim*}{Claim}

\theoremstyle{definition}
\newtheorem{definition}[theorem]{Definition}

\theoremstyle{remark}
\newtheorem{remark}[theorem]{Remark}
\newtheorem*{remark*}{Remark}

\newtheorem*{observation*}{Observation}

\newcommand{\eat}[1]{}

\newcommand{\R}{\mathbb{R}}

\newcommand{\calX}{\mathcal{X}}

\newcommand{\Esymb}{\mathbb{E}}
\newcommand{\Psymb}{\mathbb{P}}

\DeclareMathOperator*{\E}{\Esymb}

\DeclareMathOperator*{\ProbOp}{\Psymb}

\renewcommand{\Pr}{\ProbOp}

\newcommand{\eps}{\varepsilon}
\renewcommand{\epsilon}{\varepsilon}

\newif\ifnotes\notesfalse

\ifnotes
\usepackage{color}
\definecolor{mygrey}{gray}{0.50}
\newcommand{\notename}[2]{{\textcolor{blue}{\footnotesize{\bf (#1:} {#2}{\bf ) }}}}

\else

\newcommand{\notename}[2]{{}}

\fi

\title{Algorithmic Learning Foundations for Common Law\thanks{Equal contribution. This work is supported by NSF award CCF-1934931.}}

\author{Jason D. Hartline\footnote{Northwestern University, Evanston, IL, Email: hartline@northwestern.edu} \and Daniel W. Linna Jr.\footnote{Northwestern University, Chicago, IL, Email: daniel.linna@law.northwestern.edu} \and Liren Shan\footnote{Northwestern University, Evanston, IL, Email: lirenshan2023@u.northwestern.edu} \and Alex Tang\footnote{Northwestern University, Evanston, IL, Email: alextang@u.northwestern.edu}}

\begin{document}

\maketitle

\begin{abstract}
This paper looks at a common law legal system as a learning algorithm,
models specific features of legal proceedings, and asks whether this
system learns efficiently.  A particular feature of our model is
explicitly viewing various aspects of court proceedings as learning
algorithms.  This viewpoint enables directly pointing out that when
the costs of going to court are not commensurate with the benefits of
going to court, there is a failure of learning and inaccurate outcomes
will persist in cases that settle.  Specifically, cases are brought to
court at an insufficient rate.  On the other hand, when individuals
can be compelled or incentivized to bring their cases to court, the
system can learn and inaccuracy vanishes over time.
\end{abstract}

\section{Introduction}


This paper looks at a common law legal system as a learning algorithm,
models specific features of legal proceedings, and asks whether this
system learns efficiently.  Key features of our model are (a)
information is gained only by going to court, (b) going to court is
costly, and (c) individuals involved in each case decide whether or
not to go to court.  By choosing to go to court, individuals
contribute a positive informational externality to society; the
information gained can be used to inform and improve the proceedings
of future cases.  However, when court costs exceed the individual
benefits, there will be a failure of learning and inefficient
legal outcomes will persist.\footnote{Failures of learning are common in social contexts.  See \citet{GS-16} for a survey of the social learning literature.}


In a common law legal system, the law evolves as courts decide cases.
For example, the introduction of new technologies in society leads to
new factual scenarios, which require interpretation and application of
the common law.  When there is great uncertainty about how the law
would apply to these new scenarios, it is more likely that new cases
will be brought to courts.  As the legal landscape for a technology
matures, uncertainty about the governing legal principles is reduced,
and new cases are more likely to be resolved without going to court.  As the law
matures, the parties to a dispute are more likely to find precedents
(previously decided cases) that illustrate how a court would rule in
their dispute and, because of the substantial costs of going to court,
the parties will choose to resolve their dispute without bringing it to court. The resolution, given the context and legal precedents, could include the putative plaintiff walking away from a meritless claim, the putative plaintiff accepting nothing because the value of the claim is less than the expected costs of pursuing the claim, the putative defendant paying the putative plaintiff an amount to settle a claim to avoid costly litigation, or the putative plaintiff and the putative defendant agreeing to an amount to settle a claim. In this paper, we broadly refer to each of these out-of-court resolutions as a settlement, even though in some instances that parties might not have explicitly agreed to “settle.” This out-of-court resolution, however, means
that there will not be a new court ruling that contributes to refining
the applicable legal principles.


An algorithmic perspective on common law systems enables
comparative statics between a model of the status quo of legal
proceedings and opportunities for interventions.  On one extreme, the
status quo algorithm is that individuals go to court when the benefit
outweighs the cost.  On the other extreme, an algorithm can select
cases that it compels individuals to bring to court.  In between these extremes is an
algorithm that incentivizes parties to bring a case to court by offering subsidies.  Subsidies
alter the cost structure and cause, when individuals make their own
decisions, a different set of cases to go to court than would go
absent subsidies.  (In practice, subsidies can take many forms.  
For example, providing a free lawyer to an individual is a form of subsidy.)




The main features of the model in this paper are as follows.  The
outcome of a court proceeding is measured as a monetary award.  Cases
arrive online (i.e., one at a time and each case must be decided
before the next case is considered).  On arrival each case has the
option to settle (i.e., to be resolved without going to court) or to go to court.  We model the processes of
settling and going to court as running a learning algorithm on past court
data to determine an appropriate award. The difference
is that going to court discovers new relevant law (a new data
point) based on the outcome of the case while the learning algorithm of settling is only
privy to data from past court cases.  The loss function, which we aim
to minimize, is given by the squared error between the applied award
and the societally preferred award plus the costs of going to court (e.g., attorney fees, filing fees, opportunity costs, etc.).  Learning occurs
if the per-case loss vanishes as the number of cases grows; otherwise,
a gap between realized outcomes and preferred outcomes persists and
there is a failure of learning.


In this model, if the societally preferred award was known, then
settling with this award achieves a loss of zero.  The main challenge
in learning the law is that this award is not known in advance and can
only be learned by going to court.  The problem with going to court
is that it is costly.  For some types of cases (e.g., civil
infractions such as parking tickets, minor misdemeanors, small
claims), the cost of going to court can exceed the societally preferred award (or
fine, amount, etc.).  In these situations, the cost of going to court
can neutralize an otherwise efficient substantive rule that would
determine the outcome of a dispute \citep{PS-19}.  That is, these
disputes are not being decided based on the underlying substantive
law, but based on the cost of going to court.  The algorithms
considered in this paper aim to minimize both the loss from inaccurate
awards made with insufficient data and cumulative court costs.


\citet{PK-84} observed that the selection effect of which cases are
brought to court is critical for empirical understanding of court
proceedings.  Specifically, because of selection effects (e.g.,
when the parties to a dispute choose to settle), the distribution
of cases in court may bear little relation to the distribution of
disputes.  Their model was rigorously studied by \citet{KL-14} and
\citet{LK-16}.  A main contribution of our analysis is to bring to
this discussion two possibilities.  First, that the model by which
legal standards change is one of learning from past court decisions.
Second, that by compelling or incentivizing court proceedings,
selection effects are controlled.

Our model preserves a key feature of \citet{PK-84}: when the law
applicable to a dispute is well established by courts and there is
agreement about the outcome, settlement is likely.  Our model also
illustrates that some cases will not be brought to courts because of
high costs even when there is uncertainty about the outcome under 
the substantive law.  When such cases are not brought to court,
individuals, courts, and society lose a chance to learn and improve
the predictive algorithms.  With the introduction of incentives to
bring cases to court, more opportunities to learn are captured and our
model predicts more accurate legal outcomes.

The formal modeling of legal systems as learning systems, as this research develops, can
contribute to numerous strands of legal research.  For example, this
research can contribute to the long-standing debate about when law
should be fashioned into rules versus standards \citep{kap-92}.  This
debate rests on assumptions about the ex ante and ex post costs of
rules versus standards for lawmakers, courts, and individuals. It
tends to cost more to create rules ex ante, while standards are costly
for individuals to interpret and courts to apply ex post
\citep{kap-92}.  Importantly, when individuals can cheaply determine
how rules apply to contemplated acts, their conduct is more likely to
comply with those rules than with standards that are given shape into
rules after the fact \citep{kap-92}.  Thus, courts are relied upon not
only to resolve the disputes before them, but to create case law that
guides others in society.

Formal models will help us better understand the ex ante and ex post
costs of rules and standards, and will help us better understand how
these costs will change as data analytics and computation alter the
legal landscape.  In this work, we model the cost of an individual
bringing a case to court.  As this cost is imposed, a court is provided
the opportunity to further refine standards into detailed rules.

Researchers argue that rapidly increasing computational power and the
availability of data about court outcomes will fundamentally transform the
rules versus standards debate. For example, \citet{MW-14} argue that
technology will reduce the cost for individuals and lawyers to find
the law applicable to a situation, which reduces the cost of applying
standards and makes standards relatively more attractive than rules.
Going further, \citet{CN-19} argue that technological advances in data
processing and the communication of information could allow for
personalized law—specific directives for compliance with the law
communicated to individuals based on the context.  Researchers have
also explored the coding of law into algorithms, which could take
facts as inputs and produce legal decisions
\citep{liv-20}. Algorithmic systems could improve individuals'
understanding of their rights and obligations, and bring law closer to
the action, allowing individuals to tailor their behavior to comply
with the law \citep{lin-19}.  Our model is compatible with these
perspectives.  Specifically, our model allows rich features for cases
and we model both the court outcomes and settlement outcomes as the
result of learning algorithms on the available case data.

Our model is also relevant to research and practice using data
analytics to predict court outcomes. For example, \citet{ANY-16} have
developed models to make predictions in tax law, including whether an
individual is an independent contractor or employee.  An increasing
number of law firms report that they are using data analytics tools
supplied by information providers and legal technology companies to
help them choose the most advantageous court in which to litigate a
dispute \citep{EG-21}.  Predicting court outcomes is relevant for all
cases, not only high-stakes business cases, but also for cases that
impact access to justice. For example, \citet{WWAB-19} have developed
a model to try to predict when a tenant has a winnable claim against a
landlord.  Developing technologies to help individuals understand when
they have a claim can help improve access to justice for
everyone. These predictions mine past court cases, which is what our
settlement algorithm does.  Our algorithm determines whether a
case should go to court, based on such an algorithm's prediction
absent new data.  \citep{EG-21} raise the possibility that outcome prediction
tools that reduce uncertainty about outcomes in court might undermine the
flexibility of the courts to adapt to new developments.  The failure of
new discovery of law manifests in our model as the dramatic failure of
learning without incentives; while compelling or incentivizing court
appearances ensures that courts continue to adapt and learn from new cases.

Technology is also changing the costs of going to court. For example,
courts' adoption of online dispute resolution platforms can help
reduce the cost of bringing a case to court \citep{PS-19}, resulting
in more cases being resolved based on the application of substantive
law to the facts of the matter.  When cases are not brought solely
because of the cost, even when there is a high likelihood of success
given the facts and applicable law, courts and the public are deprived of
an opportunity to learn.  Additionally, reducing the costs of bringing
cases to court can lead to a more representative sample of disputes brought to
court, which can also contribute to learning.  At the same time, some
courts, such as the Civil Resolution Tribunal in British Columbia,
Canada, have significantly altered the traditional process of bringing
a dispute to court \citep{CRT-22}. These courts focus on educating
the parties about the law and leading them to a negotiated agreement
to resolve their dispute, with a litigated decision on their dispute being a
last resort.  These changes in courts lower the cost of
going to court and change the legal outcomes. Our methods can quantify
the impact of lowering the costs of going to court.

\paragraph{Related Works}
Our model formalizes the learning of a common law legal system as an online learning problem, which is widely studied in computer science literature \citep{LS-20}. There are also many previous works on online learning with strategic agents. \citet{ACHW-15} consider the online data procurement with strategic agents. They provided the algorithm to minimize the learning regret by purchasing data from agents under a budget constraint. Instead of incentivizing agents with payment, \citet{IMSW-18} incentivize exploration with selective data disclosure. They gave a policy to select the subset of history for each agent to achieve a sublinear regret.  

\paragraph{Organization}
The remainder of the paper is organized as follows.  In \Cref{s:model},
the data generating process, learning algorithms, incentives, and the
regret minimizing framework are defined.  In \Cref{s:optional}, the
situation in which individuals optionally bring their case to court is studied
and a failure of learning is observed.  In \Cref{s:compelled}, the situation in which the algorithm can compel individuals to bring their case to
court is studied.  In \Cref{s:incentivized}, the situation in which the
algorithm can incentivize individuals to bring their case to court is
studied.  In \Cref{s:individual-guarantee}, the model in which individuals
can be compelled to bring their claim to court is revisited in the stronger model in which it
is required that individuals only settle if the error from settling is
small.  In these latter sections, it is shown that errors in court
decisions vanish over time, albeit at different rates.

\label{s:intro}
\section{Model}\label{sec:model}
In this section, we describe our model for learning the law from court information.

\paragraph{Court Information}

Let $\mathcal{X}$ be the case feature space.  For a case $x \in
\mathcal{X}$ sent to the court, the court proceedings uncovers legal
information pertaining to this kind of case $y = f(x) + \eta$ , where
$f: \mathcal{X} \to [0,\alpha]$ is the unknown decision rule and $\eta
\sim \mathcal{N}(0, \sigma^2)$ is random noise.  We define an
observation $(x,y) \in (\mathcal{X} \times \mathbb{R})$ to be a pair
of the case $x \in \mathcal{X}$ and its corresponding court outcome information $y
\in \mathbb{R}$. A dataset $D \in (\mathcal{X} \times \mathbb{R})^*$
is defined as a set of observations.

\paragraph{Decision Learning Algorithms}
Judges make their decisions about specific cases by collecting new data and applying case law.  We formalize this
adaptation as a decision learning algorithm that predicts the
decision of a case based on the available dataset.  Case law is also
available to the individuals and any legal representatives, who also adapt their calculation of the settlement amounts they would accept based on a decision learning algorithm.

\begin{definition}[Decision Learning Algorithm]
Given a family of functions $\mathcal{F}$ where each $f \in \mathcal{F}$ is a function $f: \mathcal{X} \to [0,\alpha]$, a decision learning algorithm $\mathcal{L}$ is a (possibly randomized) mapping that takes a dataset $D$ as input and outputs a function $f \in \mathcal{F}$, i.e. $\mathcal{L}: (\mathcal{X} \times \mathbb{R})^* \to \mathcal{F}$. Given a dataset $D$, a decision learning algorithm $\mathcal{L}$ and a case $x \in \mathcal{X}$, we will denote the predicted court decision of $x$ as $\mathcal{L}(D)(x) \in [0,\alpha]$.
\end{definition}

For a decision learning algorithm $\mathcal{L}$ and a random dataset $D$, we define $err(\mathcal{L},D,x)$ to be an upper bound on the root-mean-square error of $\mathcal{L}$ on case $x$ based on the dataset $D$, which means 
$$
\sqrt{\mathbb{E}\bigg[(\mathcal{L}(D)(x) - f(x))^2\bigg]} \leq err(\mathcal{L},D,x).
$$
The expectation is taken over the randomness of the dataset $D = \{(x_i, y_i)\}_{i \in [m]}$ and any randomness of $\mathcal{L}$.

We define a decision learning algorithm as good if it has an error bound that decreases as $O(1/\sqrt{m})$ with the size $m$ of the dataset.

\begin{definition}\label{def:good_decision_algo}
For a true decision rule $f$, a decision learning algorithm is good if given any dataset $D = \{(x_i, y_i)\}_{i \in [m]}$ consisting of $m$ observations where  $y_i = f(x_i) + \eta_i$ and $\eta_i \sim \mathcal{N}(0, \sigma^2)$, it holds for any case $x \in \calX$:
\[
err(\mathcal{L},D,x) = O\bigg(\dfrac{\sigma}{\sqrt{m}}\bigg).
\]
\end{definition}

We consider the true decision rule $f$ is in a family of functions $\mathcal{F}$ which is learnable in expectation. For a family of functions $\mathcal{F}$, let $\mathcal{A}_\mathcal{F}$ be the set of all learning algorithms with respect to $\mathcal{F}$.

\begin{definition}[Learnability in Expectation]\label{def:Learnability}
A family of functions $\mathcal{F}$ is learnable in expectation, if for any true decision rule $f$, there exists a good decision learning algorithm $\mathcal{L}$. 
\end{definition}

Families of simple functions are known to be learnable in expectation. For example, learning the constant functions $\{\mu \in \R~|~\forall x \in \mathcal{X}, f(x)=\mu\}$ is equivalent to estimating the mean of independent Gaussian random variables $y_i \sim \mathcal{N}(\mu, \sigma^2)$. The empirical mean of $m$ samples can give a $O(\sigma/\sqrt{m})$ root-mean-square error bound. 

The family of linear functions is also learnable with an extra dependency on the dimension of the features, $n$. Consider the learning algorithm outputting the ordinary least squares (OLS) estimator of the linear function's coefficients. The following lemma gives the error bound of a learning algorithm using the OLS estimator.

\begin{lemma}\label{lem:err_bound}
Given a dataset $D = \{(x_i, y_i)\}_{i \in [m]}$ with $m$ observations where $x_i \in \{x \in \R^n : \|x\| \leq 1\}$ are i.i.d. cases and $y_i = f(x_i) + \eta_i$ with $\eta_i$ being i.i.d. noise from $\mathcal{N}(0, \sigma^2)$ and $f(x)=\beta^\top x + \beta_0$. Then there exists a learning algorithm $\mathcal{L}$ such that for any $x$, $err(\mathcal{L}, D, x) = O(\sqrt{n}\sigma/\sqrt{m})$.
\end{lemma}
\begin{proof}
Let $\beta^* = [\beta ~ \beta_0]^\top \in \R^{n+1}$, $\widetilde{x}_i = [x_i~1]^\top$, $X=[\widetilde{x}_1 ~ \widetilde{x_2} ~...~ \widetilde{x}_m]^\top$, $Y=[y_1 ~ y_2 ~...~ y_m]^\top$, and let a learning algorithm $\mathcal{L}$ output the OLS estimator of $\beta^*$. Then the output estimator $\hat{\beta}$ has the form
\begin{equation*}
    \hat{\beta} = \arg\min_{\beta \in \R^{n+1}} \|X\beta-Y\|^2 = (X^\top X)^{-1}X^\top Y
\end{equation*}
with $\E[\hat{\beta}] = \beta^*$ and $\mathrm{Cov}[\hat{\beta}] = \sigma^2(X^\top X)^{-1}$. Now, observe that
\begin{equation*}
    \E[|\mathcal{L}(D)(x) - f(x)|^2]
    \leq \E[\|\hat{\beta} - \beta^*\|^2]\E[\|x\|^2]
\end{equation*}
\begin{equation*}
    = \mathrm{trace}(\E[\mathrm{Cov}(\hat{\beta})])
    = \sigma^2 \mathrm{trace}(\E[(X^\top X)^{-1}])
\end{equation*}
Since we have assumed $\|x\| \leq 1$, by Popoviciu's inequality on variances, it can be shown the variance of every direction is bounded in $[0,1]$, and as a consequence the trace of $\E[(X^\top X)^{-1}]$ is proportionate to $O(n/m)$, which concludes the proof.
\end{proof}

\paragraph{Costs, Incentives, and Selection Algorithms}
We now describe an online learning model for learning the law. At
each time $t = 1,2,\dots, T$, a new case $x_t \in \mathcal{X}$ is
observed. The individual in this case has a cost $c_t \in
[c_{min},c_{max}]$ to bring this case to the court.  Let $d_t \in
\{0,1\}$ be the indicator variable that this case is sent to the
court, e.g. $d_t =1$ if the case $x_t$ is sent to the court; otherwise
$d_t=0$.  Then, let $D_t = \{(x_i,y_i) : d_i = 1, i \leq t\}$ be the
dataset at time $t$. If the individual chooses not to bring the case to court, then this individual will settle with outcome $\mathcal{L}(D_{t-1})(x_t)$ based on the observed dataset $D_t = D_{t-1}$. If the individual chooses to bring the case to court, then the court will uncover the noisy legal information $y_t$ and generate the court decision $\mathcal{L}(D_{t})(x_t)$ by using the decision learning algorithm $\mathcal{L}$ and the updated dataset $D_t = D_{t-1}\cup \{(x_t,y_t)\}$. We assume the court generates the court decision based on the decision learning algorithm since the court decision will be more accurate as more cases are observed. 


We will consider three models. In the first two models, we use the selection algorithm denoted by $\Pi$ to compel or incentivize individuals to go to court.
In the first model, the selection algorithm $\Pi$ can compel selected
cases to go to court.  In this case the algorithm directly selects the variable $d_t$ for each $t$.  
In the second model, the selection
algorithm $\Pi$ chooses the subsidy $s_t \geq 0$ to be paid to the
individual $t$ to defray the costs of going to court. The individual
then decides whether to go to court based on the incentive. For comparison, we also consider the third model where there is no selection algorithm. This model is equivalent to the model with subsidies where subsidies are set to zero. 

In the second model, the utility difference of this individual between going to court and settling out of court is $s_t - c_t - \mathcal{L}(D_{t})(x_t) + \mathcal{L}(D_{t-1})(x_t)$. If the individual knows the court decision $\mathcal{L}(D_{t})(x_t)$ beforehand, then the individual will prefer to go to court when this utility difference is positive. However, the court information of this case $y_t$ and the court decision $\mathcal{L}(D_{t})(x_t)$ is unknown to the individual before going to court. Thus, we assume that the individual uses the error bound of the decision learning algorithm to estimate the difference between outcomes $\mathcal{L}(D_{t})(x_t)$ and $\mathcal{L}(D_{t-1})(x_t)$. Specifically, $|\mathcal{L}(D_{t})(x_t) - \mathcal{L}(D_{t-1})(x_t)| \leq 2\,err(\mathcal{L},D_{t-1}, x_t)$. We then assume the individual knows the root-mean-square error bound $err(\mathcal{L},D_{t-1},x_t)$ and decides whether to go to court based on $c_t$, $s_t$, and error bound $err(\mathcal{L},D_{t-1},x_t)$.
If $c_t - s_t >
2\,err(\mathcal{L},D_{t-1},x_t)$, the individual chooses to settle with outcome
$\mathcal{L}(D_{t-1})(x_t)$. If $c_t - s_t \leq
2\,err(\mathcal{L},D_{t-1},x_t)$, the individual chooses to bring the
case to court. Then, the public court information $y_t = f(x_t) +
\eta_t$ is observed. The court decides the case according to
$\mathcal{L}(D_{t})(x_t)$ based on the updated dataset $D_t$. 

In the second model, we use the selection algorithm $\Pi$ to choose the subsidy to incentivize the individual to go to court. If we subsidize the individual with an amount $s_t$ that is larger than the sum of the cost of going to court $c_t$ and the court decision $\mathcal{L}(D_{t})(x_t)$, then the individual could benefit from breaking the law and going to court with the subsidy. To prevent this, we require the selection algorithm $\Pi$ to satisfy the ex ante deterrent constraint.

\begin{definition}
The selection algorithm $\Pi$ satisfies the ex ante deterrent constraint if the expected payoff for the individual to violate the law is negative at every time $t\leq T$
$$
\E_{\Pi,c_t}[s_t-c_t-\mathcal{L}(D_{t-1})(x_t)] \leq 0.
$$
\end{definition}

\paragraph{Regret}

We define the loss function $\ell_t: \mathcal{A}_\mathcal{F} \times \R \to \R$ incurred by the selection algorithm at each round $t$ with respect to the true decision rule $f$ as follows
\begin{align}\label{eqn:loss}
\ell_t(\mathcal{L}, s_t) =& (1-d_t)(\mathcal{L}(D_{t-1})(x_t) - f(x_t))^2 \nonumber\\
&+d_t((\mathcal{L}(D_{t})(x_t) - f(x_t))^2 + c_t) \nonumber \\
=& (\mathcal{L}(D_{t})(x_t) - f(x_t))^2 + d_t c_t,
\end{align}
where $d_t = \mathds{1}\{s_t \geq c_t - 2\,err(\mathcal{L},D_{t-1},x_t)\}$ denotes whether the individual accepts the settlement and the second equality is due to $\mathcal{L}(D_{t-1})(x_t) = \mathcal{L}(D_{t})(x_t)$ if $d_t = 0$. 
Note that this loss is never directly observed by the decision and selection algorithms. The decision and selection algorithms observe court information $y_t$ if and only if the individual chooses to go to court, i.e. $d_t = 1$. 

The goal of the selection algorithm is to minimize the total loss. For any selection algorithm $\Pi$, let  
$
L_\Pi = \sum_{t=1}^T \ell_t(\mathcal{L}, s_t) 
$
be the cumulative loss from $1$ to $T$. We define the regret as follows.
\begin{definition}\label{def:regret}
The regret of a selection algorithm $\Pi$ is  
\begin{align}\label{eqn:regret}
R_T = \frac{1}{T}\E_{\Pi, y_t}[L_\Pi - L^*],
\end{align}
where $L^* = \E[\sum_{i=1}^T \ell_t(\mathcal{L}^*,0)]$ is the minimum loss in the offline setting, $\mathcal{L}^*$ is the best decision learning algorithm with access to the offline dataset $\{(x_t,y_t): t=1,2,\cdots, T\}$.
\end{definition}

We note that for a family of functions $\mathcal{F}$ which is learnable in expectation, the best decision algorithm $\mathcal{L}^*$ with access to the offline dataset achieves a $O(\sigma^2/T)$ error bound. Thus, the minimum loss $L^*$ for learning $\mathcal{F}$ is $O(\sigma^2)$. As the total number of cases $T$ goes to infinity, the regret behaves asymptotically as $\frac{1}{T}\E_{\Pi, y_t}[L_\Pi]$. 



\label{s:model}
\section{Optional Court Appearances}
In this section, we consider the setting where there is no selection algorithm, i.e. all subsidies $s_t = 0$. Then, the selection is given by the incentive constraint. We show a $\Omega(1)$ lower bound on the regret for this setting. If the minimum cost $c_{min} > 0$, then the number of cases that is sent to court when there are no subsidies is a constant independent of the total number of cases $T$. As the total number of cases $T$ increases, this implies the failure of learning. 

\begin{theorem}
Suppose the error bound $err(\mathcal{L},D,x)$ of the decision learning algorithm $\mathcal{L}$ is the same across all cases $x \in \calX$. Assume the minimum cost $c_{min} > 0$. 
If there is no selection algorithm, i.e. $s_t = 0$ for all $t$, then the regret is at least $\Omega(1)$ in expectation. 
\end{theorem}

\begin{remark}
    We assume the error bound $err(\mathcal{L},D,x)$ is the same across all cases $x \in \calX$. This assumption holds when the case space is a singleton or the error bound is independent of features as in Lemma~\ref{lem:err_bound}. 
\end{remark}

\begin{proof}
Consider any decision algorithm $\mathcal{L}$ used to predict the decision. Since $s_t = 0$ for all cases $x_t$, the individual in case $x_t$ goes to court if and only if $c_t \leq 2\,err(\mathcal{L},D_{t-1},x_t)$. We note that the error bound of the decision algorithm $\mathcal{L}$ is non-increasing as the dataset $D$ increases, which means for any $D_1 \subseteq D_2$, and any case $x \in \calX$
$$
err(\mathcal{L},D_2,x) \leq err(\mathcal{L},D_1,x).
$$

For any realized sequence of $D_t$, we have two situations: (1) the error bounds are at least the minimum cost for all $t \leq T$; (2) the error bound at a time $t' < T$ is less than the minimum cost. 
In the first situation, since the error bounds satisfy $c_{min} \leq 2\,err(\mathcal{L},D_{t-1},x_t)$ for all $t \leq T$, all cases are sent to court. Thus, the expected total loss in this situation is at least $\Omega((c_{min}/2)^2 T)$.
In the second situation, there exists a time $t' < T$ such that $c_{min} > 2\,err(\mathcal{L},D_{t'-1},x_{t'})$. Since the error bound $err(\mathcal{L},D_{t'-1},x)$ is the same across all cases $x \in \calX$, the individuals in the cases after time $t'$ will have no incentive to go to court. Thus, the number of cases sent to court is always $|D_{t'}|$ which is a constant independent of the  total number of cases $T$. Given a fixed dataset, the error bound for all cases at $t > t'$ is also a fixed constant $err(\mathcal{L},D_{t'},x_t) > 0$. Since the error bound is greater than $0$, the error of each case at $t > t'$ is greater than $0$ with high probability. As the number of cases $T$ increases, the expected total loss in this situation is at least $\Omega(T)$. Combining two situations, we get the regret is at least $\Omega(1)$. \qedhere

\end{proof}
\label{s:optional}
\section{Compelled Court Appearances}
In this section, we consider the selection algorithm that compels a set of individuals to go to court. Given the number of cases $T$, we first show the sample complexity of the explore-then-commit algorithm, which compels the first $O(\alpha\sqrt{T/c_{max}})$ cases to court. When the total number of cases $T$ is unknown, we give a dynamic compelling algorithm, which compels each individual with probability $\alpha/\sqrt{tc_{max}}$ based on the time $t$, the maximum decision $\alpha$, and the maximum cost $c_{max}$. This dynamic compelling algorithm achieves the same optimal regret bound $O(\alpha\sqrt{c_{max}/T})$ as the explore-then-commit algorithm under a slightly stronger assumption on the maximum cost $c_{max}$ and the maximum decision $\alpha$. 

For each case $x_t$, we assume that the cost for the individual in this case $c_t$ is bounded in $[c_{min},c_{max}]$. The costs for individuals are unknown and adversarial. We also assume this maximum cost $c_{max}$ is known, which is taken as an input of the selection algorithm. 

We first consider the explore-then-commit algorithm, which compels the individuals in the first $m$ cases to go to court. For each case at step $t > m$, a decision algorithm $\mathcal{L}$ which satisfies the property in Definition \ref{def:good_decision_algo} is used to predict the decision. Given the number of cases $T$, we show the sample complexity of the explore-then-commit algorithm to achieve the best regret bound. 

\begin{theorem}
Suppose the decision learning algorithm $\mathcal{L}$ is good. 
Given the number of cases $T$, the maximum decision $\alpha$, the maximum cost $c_{max} \geq \alpha^2\ln^2 T/T$, the explore-then-commit algorithm compels the first $\lceil\alpha\sqrt{T/c_{max}}\rceil$ cases to go to court to achieve the best regret bound $O(\alpha\sqrt{c_{max}/T})$. 
\end{theorem}

\begin{proof}

Suppose the algorithm compels the first $m$ individuals to go to court. Since $c_t \leq c_{max}$ and the decision learning algorithm used is good, by Definition~\ref{def:good_decision_algo}, the decision error at time $t \leq m$ is bounded by $O(\sigma^2/t)$, the expected total loss for these cases is at most
\begin{align*}
\E\bigg[\sum_{t=1}^m \ell_t(\mathcal{L},s_t)\bigg] \leq mc_{\max}+\sum_{t=1}^m O\bigg(\frac{\sigma^2}{t}\bigg)\\
\leq mc_{\max}+O(\sigma^2\ln m).
\end{align*}
The remaining $T-m$ cases are directly settled by the decision algorithm $\mathcal{L}$ with an error bound $O(\sigma^2/m)$. Thus, the expected loss in these cases is 
$$
\E\bigg[\sum_{t=m+1}^T \ell_t(\mathcal{L},s_t)\bigg] = (T-m)O\bigg(\frac{\sigma^2}{m}\bigg).
$$

By taking $m = \lceil\alpha\sqrt{T/c_{max}}\rceil$, we have
\begin{align*}
    \E\bigg[\sum_{t=1}^T \ell_t(\mathcal{L},s_t)\bigg] &\leq mc_{\max}+O(\sigma^2\ln m)+(T-m)O\bigg(\frac{\sigma^2}{m}\bigg) \\
    &\leq O(\alpha\sqrt{c_{max}T}),
\end{align*}
where the last inequality is due to $\alpha \leq \sqrt{Tc_{max}}/\ln T$ and $\sigma \leq \alpha$.
\end{proof}

When the total number of cases $T$ is unknown, we use a dynamic compelling algorithm, which is also known as the epsilon-greedy algorithm \citep{LS-20}. 
For every case at time $t$, the algorithm samples a Bernoulli random variable $d_t$ with probability $p=\min\{1,\alpha/\sqrt{tc_{max}}\}$. If $d_t= 1$, the algorithm compels the individual in this case to court. Otherwise, this case is settled by a decision algorithm $\mathcal{L}$ which satisfies the property in Definition \ref{def:Learnability}. We show a regret bound of this algorithm. 

\begin{theorem}\label{thm:etc-cmax}
Suppose the decision learning algorithm $\mathcal{L}$ is good. Assume the maximum decision $\alpha \leq \sqrt{T}$ and the maximum cost $\alpha^2\ln^2 T/T \leq c_{max} \leq T\alpha^2/\ln^4T$ are given in the input. The dynamic compelling algorithm achieves a
$O(\alpha\sqrt{c_{max}/T})$ regret bound. 
\end{theorem}

\begin{proof}
If $\alpha/\sqrt{c_{max}} \geq 1$, then the individual in every case $x_t$ at time $t \leq \lfloor\alpha^2/c_{max}\rfloor$ is compelled to go to court. Let $t' = \lfloor\alpha^2/c_{max}\rfloor$. By the error bound of $\mathcal{L}$ and $\alpha \leq \sqrt{c_{max}T}/\ln T$, the expected total loss in the first $2t'$ cases is at most
\begin{align*}
    &\E\bigg[\sum_{t=1}^{2t'} \ell_t(\mathcal{L},s_t)\bigg] \leq \\
    &\leq \sum_{t=1}^{t'} O\bigg(\frac{\sigma^2}{t}\bigg) + \sum_{t=t'}^{2t'} O\bigg(\frac{\sigma^2}{t'}\bigg) + c_{max}\frac{2\alpha^2}{c_{max}} \\
    &\leq O(\alpha\sqrt{c_{max}T}).
\end{align*}

Let $m_t = |D_{t-1}|$ denote the number of cases observed in court before the case $x_t$. For each case $x_t$ at $t > 2t'$, the expected number of cases observed in court before $x_t$ is at least
\begin{align*}
\E[m_t] &= \E\bigg[\sum_{i=1}^{t-1} b_t\bigg] \geq \sum_{i=t'}^{t-1} \frac{\alpha}{\sqrt{ic_{max}}} \\
&\geq \alpha\frac{\sqrt{t}-\sqrt{t'}}{\sqrt{c_{max}}} \geq \frac{\alpha}{2}\sqrt{\frac{t}{c_{max}}}.
\end{align*}
By the Chernoff Bound, for every case at time $t \geq c_{max}(48\ln T)^2/\alpha^2$, we have
\begin{align*}
\Pr \bigg[m_t \geq \frac{1}{2}\E[m_t]\bigg] \leq \exp\bigg(-\frac{\E[m_t]}{12}\bigg) \leq \frac{1}{T^2}.
\end{align*}
Let $\mathcal{E}$ be the event that $m_t \geq \E[m_t]/2$ holds for every case at $t \geq c_{max}(48\ln T)^2/\alpha^2$. By taking the union bound over all cases at time $t \geq c_{max}(48\ln T)^2/\alpha^2$, we have the event $\mathcal{E}$ happens with probability at least $1-T$. 

For every case at time $t < c_{max}(48\ln T)^2/\alpha^2$, we upper bound the error by the maximum decision
$$
(\mathcal{L}(D_{t})(x_t) - f(x_t))^2 \leq \alpha^2.
$$ 
For every case at time $t \geq c_{max}(48\ln T)^2/\alpha^2$, the expected error for this case conditioned on the event $\mathcal{E}$ is
\begin{align*}
\E\bigg[(\mathcal{L}(D_{t})(x_t) - f(x_t))^2 \mid \mathcal{E} \bigg] &= O\bigg(\E\bigg[\frac{\sigma^2}{m_t} \mid \mathcal{E} \bigg]\bigg) \\
&=  O\bigg(\alpha \sqrt{\frac{c_{max}}{t}}\bigg),
\end{align*}
where the second equality is due to $\sigma \leq \alpha$.

If $b_t = 1$, then this agent is compelled to go to court by paying the cost $c_t \leq c_{max}$. Thus, the expected  total loss of all cases at $t > 2t'$ conditioned on event $\mathcal{E}$ is 

\begin{align*}
    &\E\bigg[\sum_{t=2t'+1}^T \ell_t(\mathcal{L},s_t) \mid \mathcal{E}\bigg] \leq \\
    &\leq \alpha^2\cdot c_{max}(48\ln T)^2/\alpha^2 + \sum_{t=1}^T O\bigg(\alpha\sqrt{\frac{c_{max}}{t}}\bigg) + c_{max}\frac{\alpha}{\sqrt{tc_{max}}} \\
    &\leq O(\alpha\sqrt{c_{max}T}),
\end{align*}
where the second inequality uses that $c_{max} \leq T\alpha^2/\ln^4T$. Combining with the bound on the first $2t'$ cases, we get the $O(\alpha\sqrt{c_{max}T})$ bound on the expected total loss.  

When the event $\mathcal{E}$ does not happen, the total error is at most $\alpha^2T$. The expected total cost is at most $O(\alpha\sqrt{c_{max}T})$. Since the event $\mathcal{E}$ does not happen with probability at most $1/T$ and $\alpha \leq \sqrt{T}$, we get the desired regret bound.
\end{proof}

\label{s:compelled}
\section{Incentivized Court Appearances}

In this section, we consider the setting where the costs for individuals to go to court $c_t$ are $i.i.d.$ drawn from an unknown distribution $P_c$, and $c_t$ is independent of the court information $y_t$.  Let $\bar{c} = \E[c_t]$ be the mean of the cost distribution. We provide a subsidy sampling algorithm that samples a random subsidy to incentivize the individual in each case. We show that this subsidy sampling algorithm improves the regret bound to $O(\sqrt{\bar{c}/T})$, while the regret bound of the compelling algorithm is $O(\bar{c}/\sqrt{T})$. If $c_{max} \leq (\bar{c}/2)^2$, we also show that this algorithm satisfies the ex ante deterrent constraint defined as follows. 

The selection algorithm $\Pi$ satisfies the ex ante deterrent constraint if at every time $t \leq T$, it holds that 
$$
\E_{\Pi,c_t}[s_t-c_t-\mathcal{L}(D_{t-1})(x_t)] \leq 0,
$$
which means the expected payoff for the individual to violate the law is negative at every time $t\leq T$.

We now describe the subsidy sampling algorithm, which is inspired by the pricing distribution in \citet{ACHW-15}. Suppose $\alpha/\sqrt{c_{min}} \leq 1$. For each case $x_t$, the algorithm chooses a subsidy $s_t$ from a distribution which satisfies that for any cost $c \in [c_{min},c_{max}]$
$$
\ProbOp\bigg[s_t \geq c - 2\,err(\mathcal{L},D_{t-1},x_t)\bigg] = \frac{\alpha}{\sqrt{t\,c}}.
$$
Let $e_t = 2\,err(\mathcal{L},D_{t-1},x_t)$. This subsidy distribution has the following density function on $[c_{min} - e_t,c_{max} - e_t]$
$$
h(x) = \frac{\alpha}{2\sqrt{t}(x+e_t)^{3/2}}. 
$$ 
This subsidy distribution also assigns a point mass of probability $\alpha/\sqrt{tc_{max}}$ at $c_{max} - e_t$. The rest probability measure is assigned to point $0$ in this subsidy distribution. 

If $\alpha/c_{min} > 1$, this subsidy distribution is not well-defined at time $t < \lfloor\alpha^2/c_{min}\rfloor$ since the total probability will be $\alpha/ \sqrt{t\,c_{min}} > 1$. For all cases at $t \leq \max\{\lfloor\alpha^2\rfloor,\lfloor\alpha^2/c_{min}\rfloor\}$, we scale the probability of subsidy distribution on $[c_{min} - e_t,c_{max} - e_t]$ by $1/\alpha$. We assume the maximum decision $\alpha \geq 1$. Then, this means we reduce the probabilities of providing subsidies with an equal proportion such that the subsidy distribution is well-defined for the first several cases. The rest probability is assigned to point $0$.

\begin{theorem}
Suppose the decision learning algorithm $\mathcal{L}$ is good. Suppose the costs $c_t$ are $i.i.d$ random variables. Assume that the maximum decision $1 \leq \alpha \leq T^{1/6}$, the mean of cost $\alpha^6/T \leq \bar{c} \leq T\alpha^2/\ln^4T$, and $c_{min} \geq \alpha^3/\sqrt{\bar{c}T}$. Given the range of cost $[c_{min}, c_{max}]$ and the maximum decision $\alpha$ in the input, the subsidy sampling algorithm achieves an $O(\alpha\sqrt{\bar{c}/T})$ regret bound. 

In particular, this algorithm satisfies the interim deterrent constraint if $c_{max} \leq (\bar{c}/2)^2$. 
\end{theorem}

\begin{proof}
If $\alpha/\sqrt{c_{min}} > 1$, then our algorithm has two phases with slightly different subsidy distributions. The transition point of our algorithm is at $t' =\max\{\lfloor\alpha^2\rfloor,\lfloor\alpha^2/c_{min}\rfloor\}$. We analyze the first phase $t\leq t'$.  

For every case at $t < 2t'$, we upper bound the error of this case by $\alpha^2$. Then, the expected total error in the first $2t'$ cases is at most $2\alpha^2t'$. When $t' = \lfloor\alpha^2\rfloor$, we have $2\alpha^2t' = O(\alpha^4) = O(\alpha\sqrt{\bar{c}T})$ since $\alpha \leq T^{1/6}$. When $t' = \lfloor\alpha^2/c_{min}\rfloor$, we have $2\alpha^2t' = O(\alpha^4/c_{min}) = O(\alpha\sqrt{\bar{c}T})$ since $c_{min} \geq \alpha^3/\sqrt{\bar{c}T}$. The expected total cost of cases sent to court in the first $t'$ cases is 
$$
\E[\sum_{t=1}^{t'} c_td_t] = \E\bigg[\sum_{t=1}^{t'} c_t\frac{1}{\sqrt{tc_t}}\bigg] =  O(\sqrt{\bar{c}T}).
$$
 
We now analyze the second phase. Let $m_t$ denote the number of cases observed in court before the case $x_t$. 
For each case $x_t$ at $t \geq 2t'$, the expected number of cases observed in court is 
$$
\E[m_t] = \E\bigg[\sum_{i=1}^{t-1} b_t\bigg] \geq \E\bigg[\sum_{i=t'}^{t-1} \frac{\alpha}{\sqrt{ic_i}} \bigg] \geq \frac{\alpha}{2}\sqrt{\frac{t}{\bar{c}}},
$$
where the last inequality is due to the convexity of function $x^{-1/2}$.

By the Chernoff Bound, we have for every $t\geq \bar{c}(48\ln T)^2/\alpha^2$
\begin{align*}
\Pr \bigg[|m_t - \E[m_t]| \geq \frac{1}{2} \E[m_t]\bigg] \leq  \exp\bigg(-\frac{\E[m_t]}{12}\bigg) \leq \frac{1}{T^2}.
\end{align*}
Let $\mathcal{E}'$ be the event that $m_t \geq \E[m_t]/2$ for every $\bar{c}(48\ln T)^2/\alpha^2 \leq t \leq T$.  By taking the union bound over all cases $t \geq \bar{c}(48\ln T)^2/\alpha^2$, we have the event $\mathcal{E}'$ with probability at least $1-1/T$. 

For every case at time $t < \bar{c}(48\ln T)^2/\alpha^2$, we upper bound the error by the maximum decision $\alpha^2$.
For every case at time $t \geq \bar{c}(48\ln T)^2/\alpha^2$, the expected error for this case conditioned on the event $\mathcal{E}'$ is
\begin{align*}
\E\bigg[(\mathcal{L}(D_{t})(x_t) - f(x_t))^2 \mid \mathcal{E} \bigg] =  O\bigg(\alpha\sqrt{\frac{\bar{c}}{t}}\bigg).
\end{align*}

If $b_t = 1$, then this agent is incentivized to go to court. Conditioned on event $\mathcal{E}'$, the expected total loss in all cases at $t\geq 2t'$ is  
\begin{align*}
    &\E\bigg[\sum_{t=2t'}^T \ell_t(\mathcal{L},s_t)\mid \mathcal{E}' \bigg] \leq \\
    &\leq \alpha^2\cdot \bar{c}(48\ln T)^2/\alpha^2 + \sum_{t=1}^T O\bigg(\alpha\sqrt{\frac{\bar{c}}{t}}\bigg) + \E\bigg[ \sum_{t=1}^T c_{t}\frac{\alpha}{\sqrt{tc_{t}}} \bigg] \\
    &\leq O(\alpha\sqrt{\bar{c}T}),
\end{align*}
where the last inequality is due to the concavity of function $\sqrt{x}$ and $\bar{c} \leq T\alpha^2/\ln^4T$. 

When the event $\mathcal{E}'$ does not happen, the total error is at most $\alpha^2T$. Since the event $\mathcal{E}'$ does not happen with probability at most $1/T$ and $\alpha \leq \sqrt{T}$, we get the conclusion.

We now show that this algorithm satisfies the ex ante  deterrent constraint if $c_{max} \leq (\bar{c}/2)^2$. For every case $x_t$, the ex ante deterrent constraint requires 
$$
\E_{\Pi,c_t}[s_t-c_t-\mathcal{L}(D_{t-1})(x_t)] \leq 0.
$$
To show this constraint is satisfied, we first compute the expected subsidy used by the algorithm. In the first phase of the algorithm, the expected subsidy for case $x_t$ at $t \leq t'$ is 
\begin{align*}
\E[s_t] &\leq \frac{1}{\sqrt{tc_{max}}}\cdot c_{max} + \int_{c_{min}}^{c_{max}} \frac{1}{2\sqrt{t}x^{3/2}} x \,\mathrm{d} x \\
&= t^{-1/2} (\sqrt{c_{max}} + \sqrt{c_{max}}-\sqrt{c_{min}}) \leq 2\sqrt{c_{max}}.
\end{align*}
Similarly, the expected subsidy in the second phase $t > t'$ is $\E[s_t] \leq 2\alpha\sqrt{c_{max}/t}$. Since $t' \geq \alpha^2$, we have $\E[s_t] \leq 2\sqrt{c_{max}}$.

Since $c_{max} \leq (\bar{c}/2)^2$ and $\mathcal{L}(D_{t-1})(x_t) \geq 0$,  we have
$$
\E_{\Pi,c_t}[s_t-c_t - \mathcal{L}(D_{t-1})(x_t)] \leq \E[s_t] - \bar{c} \leq 0. 
$$
\end{proof}

\label{s:incentivized}

\section{Individual Accuracy Guarantee}

In the previous sections, we propose models and algorithms that learn the court's true decision function assuming access to a good decision learning algorithm $\mathcal{L}$. 

However, since we only require $\mathcal{L}$ to perform well \textit{in expectation} with respect to the data distribution, it is likely that $\mathcal{L}$ produces a decision rule that makes significant mistakes on a case that occurs with low probability. Note that $\mathcal{L}$ still qualifies as a "good" learning algorithm as long as it outputs a function that makes few mistakes on most cases. This results in consequences that hinder the right of individuals who bring cases that are in general underrepresented, and in the long run may create additional adverse factors (e.g., loss of faith in the system) for bringing them to court.


Therefore, to address this issue, we explore the possibility of guaranteeing small errors on \textit{every case} the algorithm chooses to predict. 

The idea of assuring individual accuracy is first formalized in \citet{SL-07}, where it is named "knows-what-it-knows" (KWIK) learning and applied to the problem of online linear regression.

In this part, we propose an algorithm that, whenever it chooses to predict, guarantees accurate prediction with high probability by compelling at most $\widetilde{O}(n^3/\eps^4)$ cases being sent to court, where $n$ is the dimension of the case feature space. Compared to the explore-then-commit algorithm, trading off accuracy in expectation for uniform individual accuracy requires more samples by a factor of $\widetilde{O}((n/\eps)^3)$, hence our proposed algorithm yields a slower learning rate.

\paragraph{Model} We start by considering the following model:

\begin{enumerate}
    \item At time $t$, an individual with case $x_t \in \mathcal{X}$ arrives. We assume $\mathcal{X} = \{x \in \R^n: \|x\| \leq 1\}$.
    \item The algorithm decides whether to compel the current individual to send the case to court.
    \begin{itemize}
        \item If it chooses not to send the case to court, the algorithm outputs its predicted decision $\mathcal{L}(D_{t-1})(x_t)$ using past information and a learning algorithm $\mathcal{L}$ satisfying Definition \ref{def:good_decision_algo}.
        \item Otherwise the algorithm pays 1 unit of cost to observe a new information $y_t=f(x_t)+\eta_t$ and the court's decision $\mathcal{L}(D_{t})(x_t)$.
    \end{itemize}
\end{enumerate}

In this section, we also assume $\mathcal{F}$ is the family of linear functions parameterized by $\{\beta \in \R^{n+1}: \|\beta\|\leq 1\}$, with the last coordinate denoting a constant offset.

We aim to design an algorithm that satisfies the following conditions with probability at least $1-\delta$:
\begin{enumerate}
    \item If the algorithm chooses to decide a case, its prediction satisfies $|\mathcal{L}(D_{t-1})(x_t) - f(x_t)| \leq \epsilon$
    \item The number of times the algorithm sends a case to court is at most $\mathrm{poly}(n, 1/\epsilon, 1/\delta)$.
\end{enumerate}

To achieve this goal, we follow the work in \citet{SL-07} by using an online linear regression algorithm. 

\paragraph{Notations} We list the notation we use in this section below. Recall that $D_t = \{(x_i,y_i) : d_i = 1, i \leq t\}$ and $|D_t| = m$.

\begin{itemize}
    \item $X_t \in \mathbb{R}^{m \times (n+1)}$ is the data matrix with samples up to and including $t$ and the last column being a constant 1
    \item $Y_t = [y_1, ..., y_m]^\top \in \R^m$ is the public information available up to and including $t$
    \item the eigendecomposition of $X^\top X = U\Lambda U^\top$
    \begin{itemize}
        \item the $i$'th column of $U$ is the $i$'th eigenvector of $U$, $u_i$, the $i$'th diagonal element of $\Lambda$ is $\lambda_i$
        \item $r$ is the number of elements in $\Lambda$ that is at least 1
        \item $\bar{U}$ is the first $r$ columns of $U$
    \end{itemize}
    \item $\bar{q} = X\bar{U}\bar{\Lambda}^{-1}\bar{U}^\top x_t$
    \item $\bar{u} = [u_{r+1}^\top x_t, ..., u_{n}^\top x_t]$
    \item In this section, $\mathcal{L}(D_t)(x) = \hat{\beta}_t^\top x$, where $$\hat{\beta}_t = \arg\min_{\|\beta\| \leq 1} (Y_t - X_t\beta)^2$$ Note that a learning algorithm that outputs $\hat{\beta}_t$ can achieve low error, since the family of linear functions satisfies Definition \ref{def:Learnability}.
\end{itemize}

\paragraph{Algorithm} The KWIK algorithm introduced in \citet{SL-07} works as follows

\begin{enumerate}
    \item Set parameters $\alpha_1$, $\alpha_2$ at the beginning. 
    \item Receive case $x_t$
    \item If $\|\bar{q}\| \leq \alpha_1$ and $\|\bar{u}\| \leq \alpha_2$, output $\mathcal{L}(D_t)(x_t)$
    \item Otherwise compel the agent to send the case to court.
\end{enumerate}



The algorithm achieves the following performance guarantee

\begin{theorem}\label{thm:kwik}
With probability of at least $1-\delta$, whenever not sending the case to court, the algorithm guarantees $|\mathcal{L}(D_{t-1})(x_t) - \mathcal{L}(D_t)(x_t)| \leq \epsilon$ with cost at most $\widetilde{O}(\frac{n^3}{\epsilon^4})$.
\end{theorem}
\begin{proof}
By a direct application of Theorem 1 in \citet{SL-07}, we can first show that for $\alpha_1=\tilde{O}(\eps^2/{n\log n \sqrt{\log(1/\eps \delta)}})$ and $\alpha_2=\eps/4$, throughout the execution of the algorithm, whenever it chooses to predict a case, $\mathcal{L}(D_{t-1})$ is $\epsilon$-close to $f$ in probability (i.e. $\forall x, |\mathcal{L}(D_{t-1})(x) - f(x)| \leq \epsilon$ with high probability).

Now, consider if at time $t$ the algorithm can peek into the future and use $D_t$ instead of $D_{t-1}$, it will also have access to $X_t$, $Y_t$. By using $X_t$, we claim that at step (3) of the algorithm, both the norms of $\|\bar{q}\|, \|\bar{u}\|$ will be small, hence the output $\mathcal{L}(D_t)$ is also $\eps$-close to $f$ in probability.

To see why our claim is true, note that for any $x \in \mathcal{X}$, $x = X_t^\top X_t(X_t^\top X_t)^{-1}x = X_t^\top (X_t U\Lambda^{-1} U^\top x)$, implying we can determine whether $x$ is in the row space of $X_t$ by inspecting $(X_t U\Lambda^{-1} U^\top x)$. Since when $x=x_t$, it is in the $t$'th row of $X_t$, it must be in the row space of $X_t$, and by applying Lemma 13 in \citet{Aue-03} we can show that $\alpha_1$ is indeed an upper bound for $\|\bar{q}\|$ in this case, as a consequence our claim follows.

Finally, since both $\mathcal{L}(D_{t-1})$ and $\mathcal{L}(D_t)$ are $\epsilon$-close to $f$ with high probability, they must be at most $\eps' = 2\epsilon$ close to each other as well, which proves our theorem.
\end{proof}

Notice that under the KWIK setting, we require more samples than the explore-then-commit algorithm: with explore-then-commit, we always send the first $O(\sqrt{T})$ agents to court so that later prediction makes at most $O(1/\sqrt{T})$ error in expectation; in other words, to achieve an expected error $\eps$ by explore-then-commit, we must compel $O(1/\eps)$ agents to send their case to court. However, in the KWIK algorithm, to achieve an error $\eps$ for all predicted cases with high probability, we need $\widetilde{O}(n^3/\eps^4)$ cases sent to court, which presents a trade-off of a factor of $\widetilde{O}((n/\eps)^3)$.

Aside from paying unit costs for every case, we can also adopt the assumption made in the previous section that we pay an i.i.d. cost $c_t$ for sending case $x_t$ to court. If this is the case, by Hoeffding's bound on sample mean, we obtain the following corollary.  

\begin{corollary}
If the cost for sending each case to court is i.i.d. distributed according to some distribution $P_c$ with mean $c_0$, then Theorem \ref{thm:kwik} holds with cost $\widetilde{O}(c_0 \cdot n^3/\eps^4)$.
\end{corollary}
\begin{proof}
A direct application of Theorem \ref{thm:kwik} and Hoeffding's concentration inequality on the empirical mean of $c_t$ with $\widetilde{O}(n^3/\eps^4)$ samples yields the desired high probability bound. 
\end{proof}


\label{s:individual-guarantee}
\section{Conclusion}

In this paper, we model the common law legal system as a learning algorithm. The common law legal system learns from the cases observed and decided in court. From this learning algorithm viewpoint, we point out that there is a potential failure of learning when the putative plaintiff chooses an out-of-court settlement due to the cost of going to court. Through an economic analysis, we show that the legal system can learn efficiently by using a selection algorithm to compel or incentivize individuals to bring their cases to court. While these selection algorithms achieve efficient learning on average, there is no uniform guarantee for each individual case. We also provide a selection algorithm to guarantee uniform individual accuracy by compelling slightly more individuals to court. 

In our analysis, we assume that the socially agreed upon decision learning algorithm has a good error bound uniformly over all cases. Decision learning algorithms with the required error bound exist for case spaces without features and i.i.d. case samples from general feature spaces. One interesting direction is to consider case samples that are not identically distributed over a general feature space.

\label{s:conclusion}



\bibliographystyle{plainnat}
\bibliography{references}

\end{document}